\documentclass[a4paper,fleqn,usenatbib,useAMS]{mnras}

\usepackage{graphicx}	
\usepackage{amsmath}	
\usepackage{amssymb}	
\usepackage{multicol}        
\usepackage{bm}		
\usepackage{pdflscape}	

\usepackage[T1]{fontenc}
\usepackage{ae,aecompl}

\usepackage{txfonts}

\title[Turbulence and anomalous resistivity inside MCs]{Turbulence and Anomalous Resistivity inside Near-Earth Magnetic Clouds }

\author[Bhattacharjee et al.]{Debesh Bhattacharjee$^{1}$\thanks{Contact e-mail: \href{debesh.bhattacharjee@students.iiserpune.ac.in}{debesh.bhattacharjee@students.iiserpune.ac.in}} Prasad Subramanian$^{1}$\thanks{Contact e-mail: {p.subramanian@iiserpune.ac.in}} Teresa Nieves-Chinchilla$^{2}$  Angelos Vourlidas$^{3}$
\\
$^{1}$ Indian Institute of Science Education and Research, Pune\\ $^{2}$ Heliophysics science division, NASA-Goddard Space Flight Center, Greenbelt, MD (USA). \\ $^{3}$ The Johns Hopkins University Applied Physics Laboratory, Laurel MD, USA.}

\date{\today}

\pubyear{2022}

\begin{document}
\label{firstpage}
\pagerange{\pageref{firstpage}--\pageref{lastpage}}
\maketitle

\begin{abstract}
We use {\em in-situ} data from the Wind spacecraft to survey the amplitude of turbulent fluctuations in the proton density and total magnetic field inside a large sample of near-Earth magnetic clouds (MCs) associated with coronal mass ejections (CMEs) from the Sun. We find that the most probable value of the modulation index for proton density fluctuations ($\delta n_{p}/n_{p}$) inside MCs ranges from 0.13 to 0.16, while the most probable values for the modulation index of the total magnetic field  fluctuations ($\delta B/B$) range from 0.04 to 0.05. We also find that the most probable value of the Mach number fluctuations ($\delta M$) inside MCs is $\approx 0.1$. The anomalous resistivity inside near-Earth MCs arising from electron scattering due to turbulent magnetic field fluctuations exceeds the (commonly used) Spitzer resistivity by a factor of $\approx 500-1000$. The enhanced Joule heating arising from this anomalous resistivity could impact our understanding of the energetics of CME propagation.
\end{abstract}





\section{Introduction}

The solar wind is a natural laboratory for studying magnetohydrodynamic turbulence, and there is long-standing interest in issues such as the spectral properties of the turbulent cascade, turbulent dissipation and more. A review of early work can be found in \cite{1995GoldsteinARA&A} and a somewhat more recent review is \cite{2013BrunoLRSP}. Here we focus instead, on the interplanetary counterparts of solar coronal mass ejections (ICMEs). While the large-scale structure of ICMEs has been extensively studied, relatively less attention has been paid to turbulent fluctuations inside them.
\cite{1997RuzmaikinJGR} studied the magnetic energy spectra and helicity distribution inside 6 near-Earth MCs which were shortlisted by \cite{1990LeppingJGR} using ISEE3 satellite data. They found that the low-frequency spectrum of magnetic energy differs significantly from that of the solar wind from coronal holes. This suggests that the nature of turbulence arising from closed coronal regions (e.g.; inside MCs) is very different from the solar wind turbulence from coronal holes. A detailed study of Alfv\'enic fluctuations inside an MC detected by the Wind spacecraft at 1 AU shows that the power spectral density of magnetic fluctuations inside the MC are more anisotropic than those in the ambient solar wind \citep{2012LiangPlST}. \cite{2017LiApJ} investigated Alfv\'enic fluctuations inside ICMEs using Voyager 2 observations between 1 and 6 AU. Their findings suggest that Alfv\'enic fluctuations can contribute significantly to the local heating of the ICME plasma. While these studies concentrate on the magnetic field fluctuations inside ICMEs close to or beyond the Earth, \cite{2002LynchGeoRL} address the density fluctuations inside a CME at 10 $R_{\sun}$. They studied the interplanetary scintilation (IPS) of the radio source 0854+201 at 8 GHz during the passage of a CME along the line of sight. By simulating the observations obtained from the LASCO (Large Angle and Spectrometric Coronagraph) C3 coronagraph, they found that the electron density modulation index ($\delta n_e / n_e $) shows a 19\% drop inside the CME as compared to the pre-CME solar wind. Turbulence in the ICME sheath has been a subject of recent interest. \cite{2020GoodEGUGA} studied the evolution of magnetic field turbulence in the sheath of an ICME that encountered MESSESNGER spacecraft at 0.47 AU and subsequently STEREO-B
at 1.08 AU while the spacecraft were close to radial alignment. \cite{2021KilpuaFrASS} reported an extensive statistical analysis of the magnetic field fluctuations of 79 well-observed ICME-driven sheath regions using the high-resolution magnetic field investigation (MFI) data from the Wind spacecraft. They found that the characteristics of ICME-driven sheaths significantly differ from planetary magnetosheaths and that ICME shocks do not reset the solar wind turbulence, as is the case downstream of a bow shock. Magnetic turbulence in the ICME-driven shock was also studied by \cite{2021ZhaoA&A} where they tracked the ICME at 0.8 AU and 1 AU using the data from the Solar Orbiter and Wind spacecraft. They found an enhancement in the power in the magnetic field fluctuations downstream of the shock. They also found that the frequency of the oblique kinetic Alfv\'en waves in the downstream region is increased by a factor of $\approx 7-10$ times as compared to the upstream region.

However, we are not aware of a large-sample study of fluctuations in the plasma properties inside near-Earth ICMEs or their associated magnetic clouds (MCs) using \textit{in-situ} data. 
The rest of the paper is organized as follows: we first outline the statistics of the density, (total) magnetic field and acoustic Mach number fluctuations inside a large sample of well-observed near-Earth MCs detected by the Wind satellite (\url{https://wind.nasa.gov/}). We then calculate the anomalous resistivity inside the MCs arising from the turbulent fluctuations. The data used in this paper are described in Section~\ref{S - Data} and the events are listed in Table~\ref{S - Table A}. In Section~\ref{S - Fluctuations} we survey the amplitude of density and magnetic fluctuations inside the 
MCs and compute the acoustic Mach number associated with these fluctuations. In Section~\ref{S - anomalous resistivity}, we formulate an anomalous resistivity mechanism arising from the enhanced particle scattering caused by magnetic fluctuations inside MCs. This could result in enhanced plasma heating and increase our understanding of CME dynamics as they propagate through the heliosphere. The overall summary of this work is presented in Section~\ref{S - conclusions}.

\section{Data}
\label{S - Data}
We use {\em in-situ} data from the Wind spacecraft for this study. The Wind ICME catalogue (\url{https://wind.nasa.gov/ICMEindex.php}) lists a sample of well observed Earth directed ICMEs \citep{18NCSo, 2019NCSoPh} at the position of the Wind spacecraft. In this paper, we limit our study to MCs, which are the magnetically well-structured parts of ICMEs, with typically better defined boundaries and expansion speeds \citep{1982KleinBurlagaJGR}. The MCs associated with these ICMEs are classified into different groups depending upon how well the observed magnetic and plasma parameters fit the expectations of a static flux rope configuration. Of the ICMEs observed between 1995 and 2015 listed on the Wind website, we first shortlist MCs that are categorized as F+ and Fr events. These events best fit the expectations of the flux rope model described in \citet{2016NCApJ, 18NCSo}. Fr events indicate MCs with a single magnetic field rotation between $90^{\circ}$ and $180^{\circ}$ and F+ events indicate MCs with a single magnetic field rotation greater than $180^{\circ}$. 
We further shortlist events that are neither preceded nor followed by any other ICMEs or ejecta within a window of two days ahead of and after the event under consideration. This helps us exclude possibly interacting events from our dataset. We also identify a stretch of ambient/background solar wind for each ICME in our sample. The background is a 24-hour window in the 5 days preceding the event and satisfying the following conditions: a) the rms fluctuations of the solar wind velocity for this 24-hour period should not exceed 10\% of the mean value b) the rms fluctuations of the total magnetic field for this 24-hour period should not exceed 20\% of the mean value c) there are no magnetic field rotations. The first two criteria ensure that the chosen background is quiet. Criterion c) distinguishes the background from the MCs, because MCs are characterized by large rotations of magnetic field components and low plasma beta. We find that the average plasma beta in the background is at least 1.5 times higher than that in the MC. Our final shortlist comprises 152 near-Earth ICMEs, which are listed in Table \ref{S - Table A} of the appendix. 

We are interested in the amplitude of turbulent fluctuations inside near-Earth MCs. However, fluctuation amplitudes make sense only when compared with a suitable average value. The modulation index, which is the ratio of the fluctuation amplitude to the average, gives an idea of the ``strength'' of the turbulent fluctuations. The choice of an average value is subjective; one could compute a temporal average value (the average density, for instance) over the entire MC, or divide the MC into several (temporal) boxes and compute the moving average based on the size of these boxes. We choose the latter option, and compare the rms fluctuation amplitude inside each running box with the moving average within that box. The Wind data which we use in this study have a cadence of $\approx$ 1 minute. We compute rms fluctuations of the density and (total) magnetic field using running box averages with two temporal boxes: $t_{box}$ = 40 minutes and 60 minutes. Using $t_{box} = 40$ minutes yields $\approx$ 38 boxes inside a typical MC in our sample and $t_{box} = 60$ minutes yields $\approx$ 25 boxes inside a typical MC. By comparison, \citet{1995BavassanoJGR} use $t_{box}$ = 45, 90 and 180 minutes and \citet{2004SpanglerPhPl} use $t_{box}$ = 60 minutes in their studies of solar wind turbulence. We use exponentially weighted moving averages (EWMA) of our data.

The EWMA of a time series $\tilde{X}(i)$ is defined as
{\begin{eqnarray}
\nonumber
	X(i) = w \times \tilde{X}(i) + (1-w) \times X(i - 1) \, \, , \,  i > 1 \\
X(i) = \tilde{X}(i) \, \, , \, i = 1 \, .
\label{eq: movavg}	
\end{eqnarray}}
where $X(i)$ is the $i^{th}$ timestamp of the EWMA of $\tilde{X}(i)$ and $w$ is the weight ($0 \leq w \leq 1$). We use $w = 1/l$, following \citet{1986Hunter}. The quantity $l$ denotes the number of data points inside a given $t_{box}$ (so a box with $t_{box} = 45$ minutes would have $l = 45$, for instance) 



The quantity $\tilde{X}$ is the original one minute resolution data (of the proton density or the total magnetic field) and $X$ is the exponentially weighted moving box average data. $i$ is an integer which runs from 1 to the length of the entire dataset. Since the weights decline exponentially as the data points get older, the EWMA emphasizes newer data and better captures small variations than a simple moving average which applies equal weights to all data points \citep{1990Lucas}.

We also calculate a quantity $\delta X$ which is the rms deviation of $\tilde{X}$ from $X$ for a given $t_{box}$. We use this to form a time series of the modulation index $\delta X/X$. We then calculate the temporal average of the modulation index inside the magnetic cloud and the ambient (background) solar wind as follows:

{\begin{eqnarray}
\nonumber
\bigg \langle \frac{\delta X}{X} \bigg \rangle_{MC} \equiv \frac{1}{T_{MC}}\, \int_{{t_s}} ^{{t_e}} \frac{\delta X}{X} dt \, \\
\bigg \langle \frac{\delta X}{X} \bigg \rangle_{BG} \equiv \frac{1}{T_{BG}}\, \int_{{BG\,s}} ^{{BG\,e}} \frac{\delta X}{X} dt \, ,
\label{eq: avg}
\end{eqnarray}}
where $T_{MC}$ and $T_{BG}$ are the durations of MC and background respectively. The quantities $t_s$ and $t_e$ are the start and end times of each magnetic cloud \citep[see Figure 1 of ][]{2022DebeshSoPh} and $BG\,s$ and $BG\,e$ are the start and end times of the corresponding background solar wind.

\section{Density and Magnetic field Fluctuations}
\label{S - Fluctuations}
The \textit{in-situ} data from the Wind spacecraft provides us the temporal profiles of plasma velocity ($v$), proton number density ($n_p$), total magnetic field ($B$) and the proton thermal velocity ($v_{th_{p}}$) in the spacecraft frame of reference. We first concentrate on proton density ($n_{p}$) and total magnetic field ($B$) fluctuations by using $X \rightarrow n_{p}$ and $X \rightarrow B$ in Equations~\ref{eq: movavg} and \ref{eq: avg}. We examine the behavior of $\delta n_{p}/n_{p}$ and $\delta B/B$ inside near-Earth MCs and in the background solar wind. These quantify the extent to which the density and magnetic field are modulated due to turbulent fluctuations and give us an overall idea about the strength of density and magnetic turbulence. Using these quantities, we also quantify the magnetic compressibility of the turbulent fluctuations and the Mach number of the fluctuations. We use the term `solar wind' only for the ambient/background solar wind throughout this paper.

\subsection{The density ($\delta n_{p}/n_{p}$) and magnetic field ($\delta B/B$) modulation index}
\label{S - deltaBovB}
\begin{figure}   
                                \centering
  \includegraphics[width=\hsize]{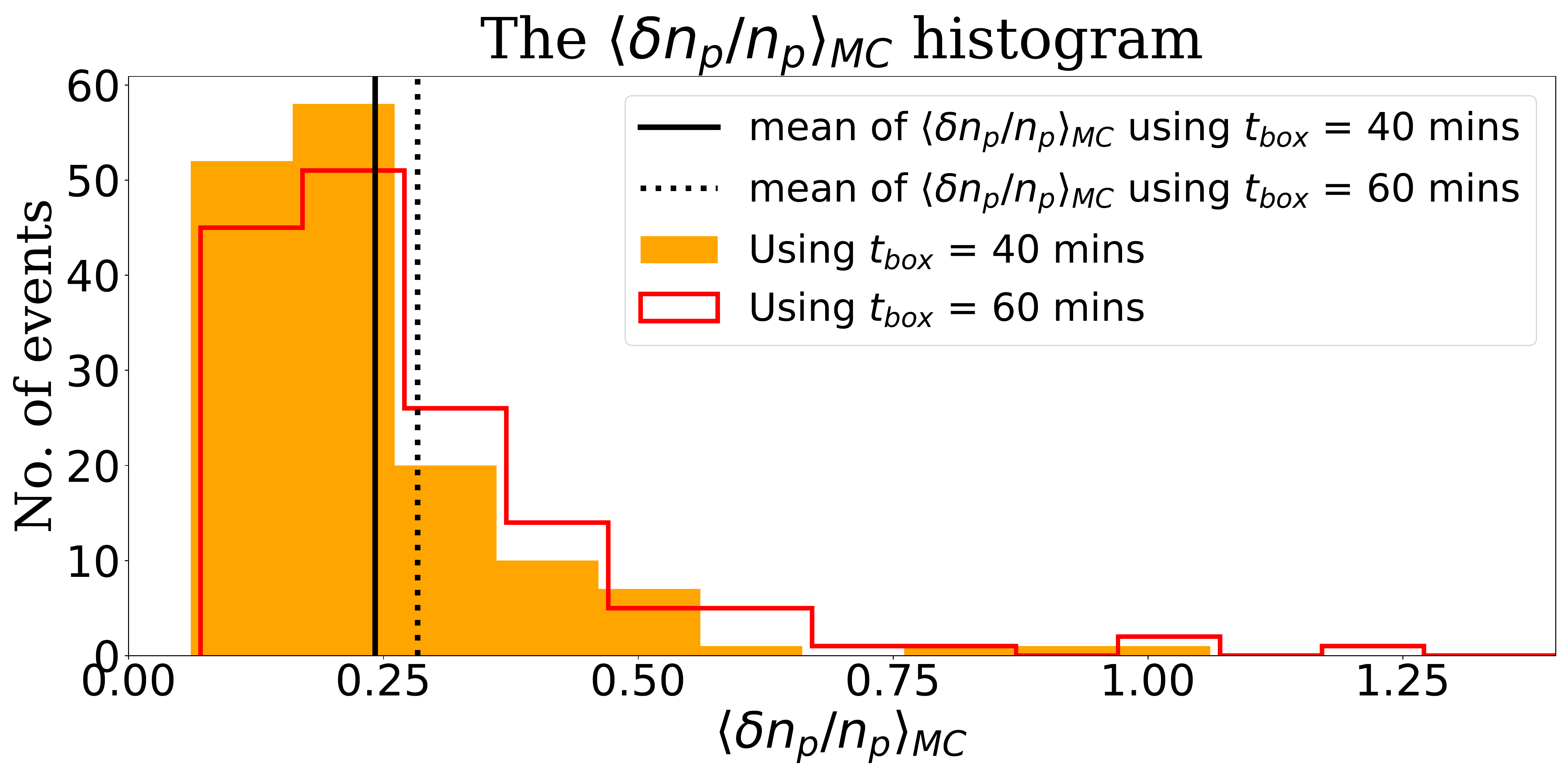}

\caption{Histograms of the proton density modulation index inside the MCs ($\langle \delta n_{p}/n_{p} \rangle_{MC}$) for all the events listed in Table~\ref{S - Table A} for the two averaging intervals. The vertical lines mark the mean values. With $t_{box} = 40$ minutes the mean, median and most probable value of $\langle \delta n_{p}/n_{p} \rangle_{MC}$ are 0.24, 0.2 and 0.13 respectively. The corresponding values for $t_{box} = 60$ minutes are 0.28, 0.23 and 0.16 respectively.}
   \label{Figure - delta_n_mc}
   \end{figure}

Figure~\ref{Figure - delta_n_mc} shows histograms of the average proton density modulation index inside near-Earth MCs ($\langle \delta n_{p}/n_{p} \rangle_{MC}$) calculated following Equations~\ref{eq: movavg} and \ref{eq: avg} for all the events listed in Table~\ref{S - Table A}. The mean, median and the most probable value of $\langle \delta n_{p}/n_{p} \rangle_{MC}$ for $t_{box}$ = 40 minutes are 0.24, 0.2 and 0.13 respectively; while for $t_{box}$ = 60 minutes, they are 0.28, 0.23 and 0.16 respectively. 
The mean, median and most probable value of the average proton density modulation index in the background solar wind ($\langle \delta n_p/n_p \rangle_{BG}$) are 0.11, 0.1 and 0.08 respectively for $t_{box}$ = 40 minutes and are 0.12, 0.11 and 0.09 respectively for $t_{box}$ = 60 minutes. 
Our results for $\langle \delta n_p/n_p \rangle_{BG}$ are in broad agreement with the findings of \cite{1995BavassanoJGR} who found that the density modulation index is a little less than 0.1 in the quiet solar wind, suggesting that the fluctuations are nearly incompressible. 

\begin{figure}   
                                \centering
  \includegraphics[width=\hsize]{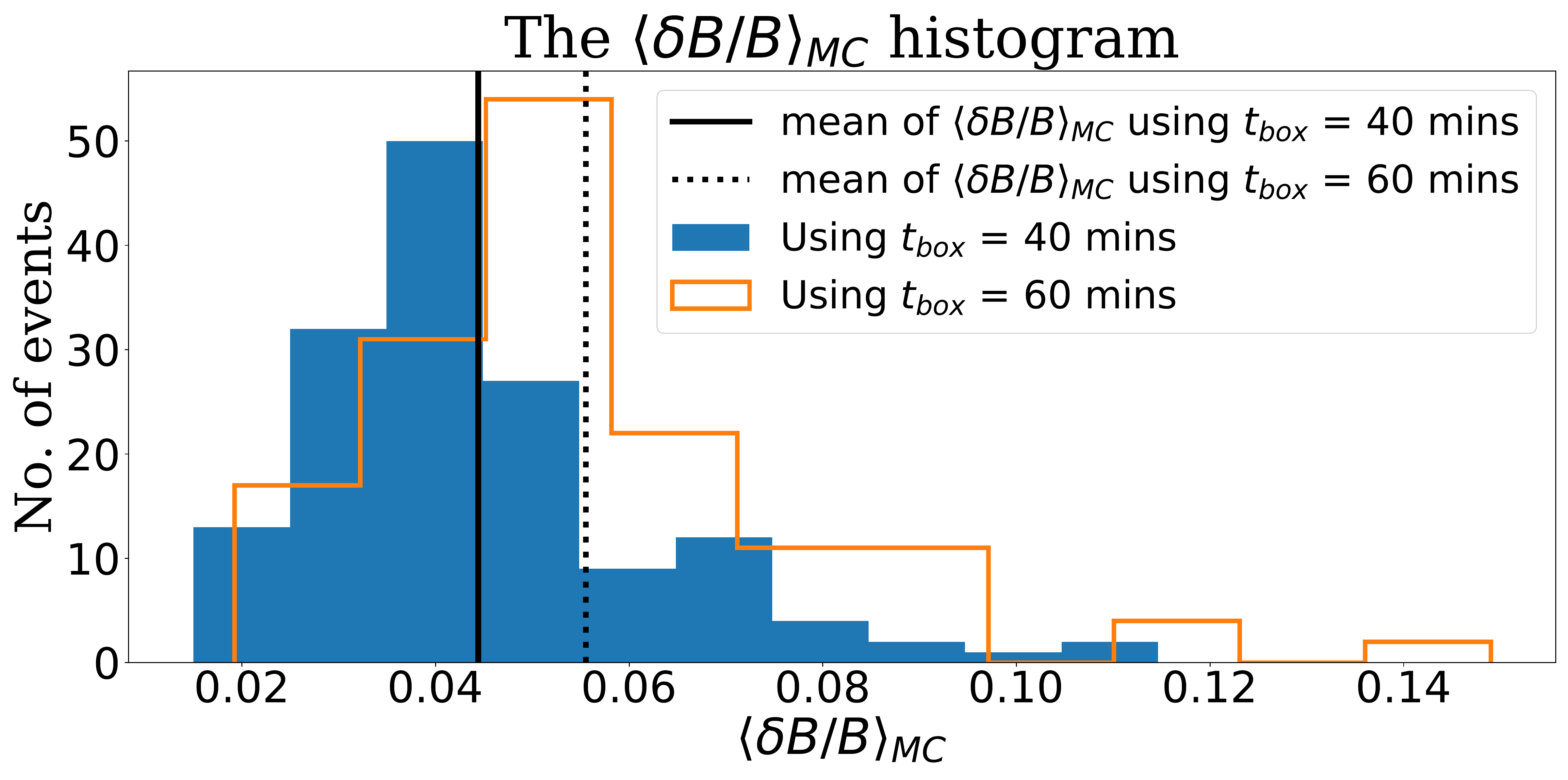}

\caption{Histograms of the total magnetic field modulation index inside the MCs ($\langle \delta B/B \rangle_{MC}$) for all the events listed in Table~\ref{S - Table A}. The vertical lines mark the mean values. With $t_{box} = 40$ minutes the mean, median and most probable value of $\langle \delta B/B \rangle_{MC}$ are 0.044, 0.042 and 0.038 respectively. The corresponding values for $t_{box} = 60$ minutes are 0.056, 0.052 and 0.05 respectively.}
   \label{Figure - delta_B_mc}
   \end{figure}

Histograms of the average magnetic field modulation index inside magnetic clouds are shown in Figure~\ref{Figure - delta_B_mc}. The mean, median and most probable value of $\langle \delta B/B \rangle_{MC}$ with $t_{box} = 40$ minutes are 0.044, 0.042 and 0.038 respectively, and for $t_{box}$ = 60 minutes, they are 0.056, 0.052 and 0.05 respectively. 
The mean, median and most probable value of the temporal average of the magnetic field modulation index in the background solar wind ($\langle \delta B/B \rangle_{BG}$) are 0.067, 0.06 and 0.059 respectively for $t_{box}$ = 40 minutes and 0.074, 0.068 and 0.063 respectively for $t_{box} =$60 minutes. We find that the magnetic field modulation index in the background solar wind ($\langle \delta B/B \rangle_{BG}$) is slightly larger than the modulation index ($\langle \delta B/B \rangle_{MC}$) inside near-Earth MCs. 
All the results regarding density and magnetic field fluctuations are listed in Table~\ref{S - Table B} of the appendix.


Interestingly, a comparison of Figures~\ref{Figure - delta_n_mc} and \ref{Figure - delta_B_mc} shows that the density modulation index inside MCs is almost an order of magnitide larger than the magnetic field modulation index. We also find that the proton density modulation index is larger than the magnetic field modulation index in the background solar wind, although to a lesser extent. This suggests that the magnetic field fluctuations are fairly compressible, and are likely due to magnetosonic modes (as opposed to Alfv\'en modes, which are incompressible.)

\subsection{Mach number fluctuations ($\delta M$)}

The fluctuation in the acoustic Mach number is defined as \citep{1995BavassanoJGR}
\begin{equation}
\delta M \equiv \frac{\delta v}{c_{s}} \, ,
\end{equation}
where $\delta v$ is the rms fluctuation in the plasma velocity and $c_{s}$ is the (moving-average) adiabatic sound speed. The adiabatic sound speed is defined as \citep{1995BavassanoJGR} $c_{s} \equiv [\gamma (n_{e}k_{B}T_{e} + n_{p} k_{B} T_{p})/(n_{p} m_{p} + n_p m_e)]^{1/2}$
where $n_{e}$ and $n_{p}$ are the electron and proton number densities in cm$^{-3}$ respectively, $T_{e}$ and $T_{p}$ are the electron and proton temperatures in K respectively, $k_{B}$ is the Boltzmann constant, $m_{p}$ and $m_e$ are the proton and electron masses in g respectively, $\gamma$ is the polytropic index and we have assumed $n_{e} = n_{p}$ and $T_{e} = 10 T_{p}$ \citep{2022DebeshSoPh}. The expression for $c_s$ uses a polytropic index $\gamma$. The appropriate value to use for $\gamma$ inside ICMEs or MCs is not clear. A value of 5/3 would imply that the ICME plasma is cooling adiabatically and needs to be heated continuously to maintain its temperature \citep{1996KumarRustJGR}. On the other hand, a value of 1.2 implies efficient thermal conduction to the interior of the ICME from the solar corona and little additional heating \citep{1996ChenJGR}. A recent study using Helios and Parker Solar Probe (PSP) data postulates a polytropic index ranging from 1.35 to 1.57 for solar wind protons and an index ranging from 1.21 to 1.29 for the solar wind electrons \citep{2022Dakeyo}. Another study which uses PSP data \citep{2020NicolaouApJ} claims a polytropic index $\approx 5/3$ for the solar wind plasma. The polytropic index of the CME plasma is thought to be in the range from 1.35 to 1.8 \citep{2018MishraApJ}. Considering this view, we use two values for $\gamma$ (5/3 and 1.2) in our calculations. 

\begin{figure}   
	\centering
	\includegraphics[width=\hsize]{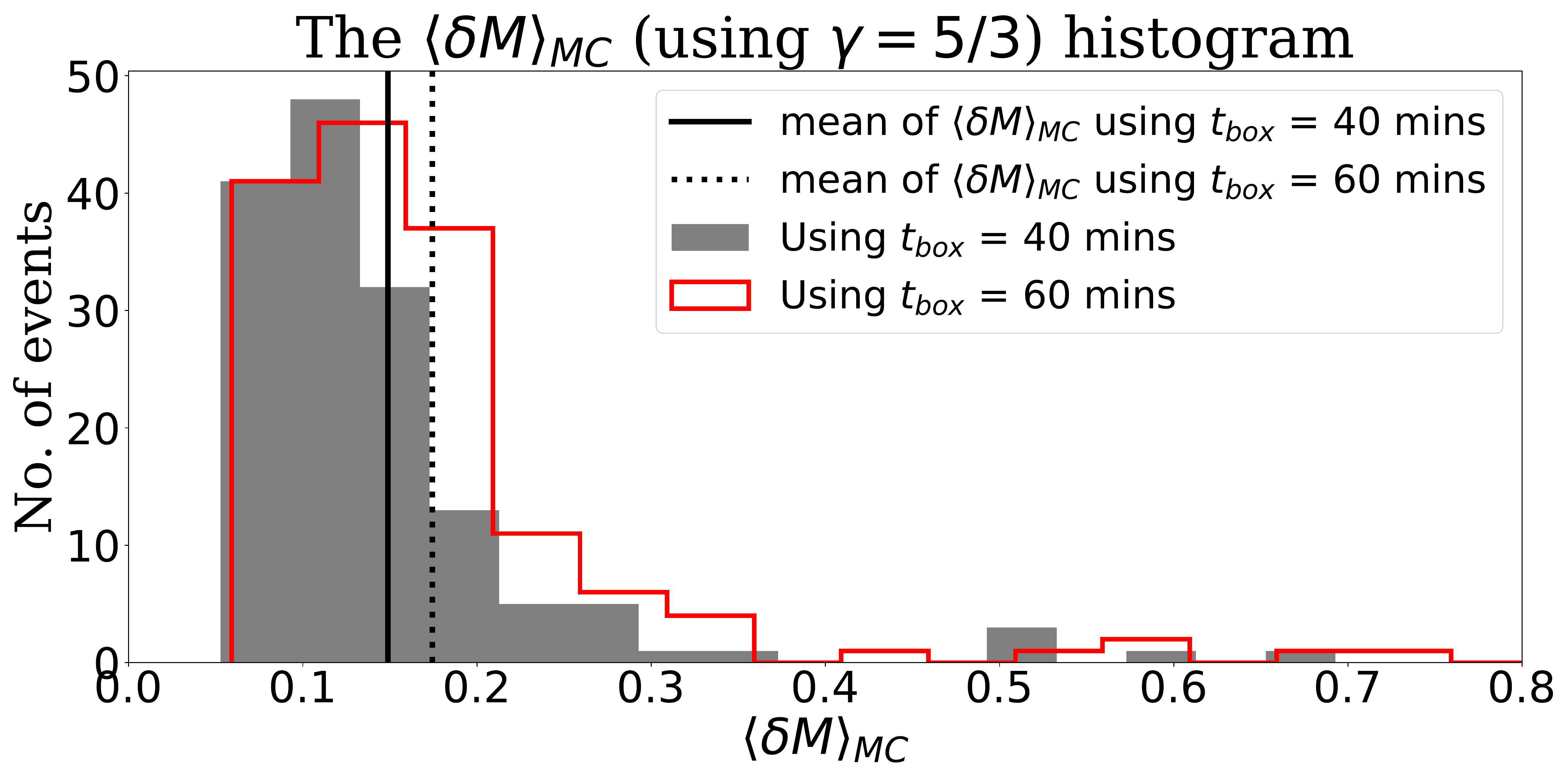}

	\caption{Histograms of the Mach number fluctuations inside the MCs ($\langle \delta M \rangle_{MC}$) using $\gamma = 5/3$ for all the events listed in Table~\ref{S - Table A}. The vertical lines mark the mean values. With $t_{box} = 40$ minutes the mean, median and most probable value of $\langle \delta M \rangle_{MC}$ are 0.15, 0.12 and 0.09 respectively. The corresponding values for $t_{box} = 60$ minutes are 0.17, 0.14 and 0.10 respectively.}
	\label{Figure - deltaM_mc}
\end{figure}

The mean, median and most probable value of Mach number fluctuations in the background solar wind $\langle \delta M \rangle_{BG}$ using $\gamma = 5/3$ for $t_{box} =$ 40 minutes are 0.84, 0.46 and 0.12 respectively while for $t_{box} =$ 60 minutes they are are 0.95, 0.52 and 0.14 respectively. Using $\gamma = 1.2$, the mean, median and most probable values become 0.98, 0.54 and 0.14 respectively (for $t_{box} = 40$ minutes) and 1.09, 0.60 and 0.15 respectively (for $t_{box} = 60$ minutes). 
Figure~\ref{Figure - deltaM_mc} shows histograms of $\langle \delta M \rangle_{MC}$ inside the MCs using $\gamma = 5/3$ for all the events listed in Table~\ref{S - Table A} for two averaging time windows. The mean, median and the most probable value of $\langle \delta M \rangle_{MC}$ for $t_{box} = $ 40 minutes are 0.15, 0.12 and 0.09 respectively. For $t_{box} = $ 60 minutes, they are 0.17, 0.14 and 0.10 respectively. If we use $\gamma = 1.2$ instead of 5/3, the mean, median and the most probable values are 0.17, 0.14 and 0.11 respectively (for $t_{box} = 40$ minutes) and 0.21, 0.16 and 0.12 respectively (for $t_{box} = 60$ minutes). We thus find that the statistical behavior of $\langle \delta M \rangle$ does not depend significantly on the choice of the polytropic index $\gamma$.
Since the mean, median and mode $\langle \delta M \rangle_{MC}$ are all considerably lesser than 1 (Figure~\ref{Figure - deltaM_mc}) the turbulent plasma velocity fluctuations inside near-Earth MCs are clearly subsonic; on the other hand, the bulk motion of ICMEs is well known to be supersonic. This leads us to consider Morokovin's hypothesis, which suggests that subsonic laws are adequate to describe the essential dynamics (such as aerodynamic drag, for instance) of an object moving at moderately supersonic speeds, as long as the turbulent fluctuations in the boundary layer are subsonic \citep{1996Smits, 2011DuanJFM}. \cite{2003WeiJGRA,2003WeiGeoRL} estimate boundary layer widths using 80 well-observed MCs from 1996 to 2001. They show that the average thickness of the front boundary layer of the MC is $\approx 1.7$ hours after the `MC start' (inside the cloud) while the average thickness of the trailing boundary layer is $\approx 3.1$ hours after the `MC end' (outside the cloud). Using $t_{box} = 10$ minutes (so as to get an adequate number of data points inside the boundary layer) and $\gamma = 5/3$, we find that the mean, median and most probable value of $\langle \delta M \rangle$ inside the front boundary layer are 0.14, 0.07 and 0.05 respectively while they are 0.16, 0.05 and 0.04 respectively in the trailing boundary layer. If we assume $\gamma = 1.2$ (instead of 5/3), the mean, median and most probable value become 0.17, 0.08 and 0.05 respectively in the front layer and 0.18, 0.06 and 0.04 respectively in the trailing layer. The choice of $\gamma$ does not affect the results substantially. The statistics for $\langle \delta M \rangle$ inside MCs, backgrounds and boundary layers are listed in Table~\ref{S - Table B}. 
Interestingly, the commonly used aerodynamic drag laws for CME propagation (both for the cases, when the drag force is proportional to the square of the relative speed between the CME and the solar wind ($V_{rel}$), \citep{2004CargillSoPh,2009Vrsnak,2010Vrsnak,2013Vrsnak} and when the drag force is proportional to $V_{rel}$ \citep{2002Vrsnak, 2009BorgazziAA}) are based on experimental data for subsonic propagation. Our findings on the subsonic nature of $\delta M$ in MC boundary layers could possibly explain why these subsonic aerodynamic drag laws work well for CMEs, even though their bulk motion is supersonic.

\section{Anomalous resistivity inside MCs}
\label{S - anomalous resistivity}

We next ask if the turbulent fluctuations inside magnetic clouds can contribute to an anomalous resistivity and associated Joule heating. It is well known that some source of heating is needed to account for the observed proton temperature profile in the solar wind \citep{1995VermaJGR}. Specifically, if magnetic clouds associated with ICMEs were to expand (and consequently cool) adiabatically without any additional heating, the proton temperature inside them would have been as low as a few Kelvin near the Earth \citep{1993CGGeoRL}, whereas the observed proton temperatures are $ \approx 10^{5}$ K. Although we have only mentioned proton heating, we note that semi-analytical treatments \citep{1996ChenJGR,1996KumarRustJGR} and MHD simulations \citep{2008ManchesterApJ,2013LionelloApJ,2017JinApJ} of CME dynamics and propagation assume a single species fluid.  Energy dissipation in such treatments, which is often parametrized using a polytropic index, refers to dissipation in the ``plasma'', without distinguishing between protons and electrons. In particular, studies which infer plasma heating rates using observations \citep{2001CiaravellaApJ,2007RakowskiApJ}  implicitly refer to electron heating. While some treatments invoke turbulent dissipation to account for proton heating \citep[e.g.,][]{2006LiuJGRA}, it is not clear if indirect electron heating via Coulomb coupling between protons and electrons is adequate. Joule (often called Ohmic) heating is a ready candidate for direct electron heating; on account of their lower mass, electrons are the current carriers, and will be the primary recepients of Joule heating. However, Joule heating due to Spitzer resistivity (which is only due to Coulomb collisions), is well known to be quite inadequate \citep{2011MurphyApJ}. We therefore focus on an anomalous (i.e., non-Spitzer) resistivity that can account for direct electron heating via Joule dissipation.

MHD turbulence is typically characterized by small scale fluctuations riding on a large-scale background magnetic field. Charged particles (we concentrate here on electrons) which nominally gyrate around the background magnetic field (and are thus ``tied'' to them) are stochastically scattered by the turbulent fluctuations in the magnetic field. This can be thought of as electrons undergoing collisions with magnetic scattering centers, thus undergoing diffusion parallel to and perpendicular to the background magnetic field. We use the results of \cite{2004CandiaJCAP}, who report extensive simulations of charged particle dynamics in the presence of turbulence, and give analytical fitting formulae for the particle diffusion coefficients due to turbulent fluctuations. Their results are consistent with earlier ones reported by \cite{1999GiacaloneApJ, 2001Casse} and are also valid for higher energy particles and stronger turbulence levels. Although these particle diffusion coefficients are normally used for high energy cosmic rays, they are technically valid for charged particles of any energy, and are quite applicable to the situation we consider. We use our estimates of the turbulence amplitudes inside near-Earth MCs (Section~\ref{S - deltaBovB}) in these fitting formulae for diffusion coefficients and extract effective (anomalous) collision timescales. These timescales can be used (instead of the usual Spitzer/Coulomb collision timescales) to compute an anomalous resistivity.

\cite{2004CandiaJCAP} give the following fitting formula for the spatial diffusion coefficient (${\rm cm^{2}\,s^{-1}}$) parallel to the large-scale background magnetic field: 

\begin{equation}
D_\parallel = v_{th_{e}} L_{max} \rho \times \frac{N_\parallel}{(\delta B / B)^2} \times \sqrt{\left(\frac{\rho}{\rho_\parallel}\right)^{2(1-\Gamma)} + \left(\frac{\rho}{\rho_\parallel}\right)^2}
\label{eq: D_para}
\end{equation}
where $v_{th_{e}}$ is the electron thermal velocity and $L_{max}$ is the largest turbulent length scale of the system. We note that the corresponding formula in \cite{2004CandiaJCAP} uses the speed of light ($c$) in place of the thermal velocity $v_{th_{e}}$, since their treatment addresses relativistic particles. The quantity $\delta B/ B$ is the modulation index of magnetic fluctuations we have discussed in Section~\ref{S - deltaBovB}, and gives a measure of the strength of turbulent fluctuations. The quantity $\rho$ is the ratio of the electron gyroradius ($r_L$) to $L_{max}$ and it characterizes how tightly the particle is tied to the background magnetic field. The electron gyroradius ($r_L$) is given by \citep{2013NRLHuba},
\begin{equation}
r_L = 2.38 \times {T_{eV}}^{1/2} \times B^{-1} \, {\rm cm}
\label{eq: rL}
\end{equation}
where $T_{eV}$ is the electron temperature in electronvolt (eV) and $B$ is the total magnetic field in Gauss. 
Common choices for the maximum lengthscale $L_{max}$ include the MC width ($L_{MC}$) \citep{2013ArunBabu} and the outer scale of solar wind turbulence, which is $\approx 10^{6}$ km \citep{2009Prasad}. We use both these prescriptions, starting with $L_{max} = L_{MC}$. The quantity $L_{MC}$ is calculated as \citep{2022DebeshSoPh}
\begin{equation}
L_{max} = L_{MC} \equiv \int_{t_{\rm s}}^{t_{\rm e}}   |{v}(t)| dt \, .
\label{eq: Lmc}
\end{equation}  
where $v$ is the plasma velocity measured in the spacecraft reference frame, $t_s$ and $t_e$ denote the start and end time of the MC \citep[see Figure 1 of ][]{2022DebeshSoPh}. The quantities $\Gamma$, $N_{\parallel}$ and $\rho_{\parallel}$ are fitting parameters which depend on the type of the turbulent energy spectra we would like to consider (Kolmogorov, Kraichnan or Bykov-Toptygin). Since the energy spectrum for magnetic fluctuations in MHD turbulence follows the Kraichnan scaling $E(k) \propto k^{-3/2}$ \citep{1993Biskamp} and because several solar wind turbulence observations show evidence for this \citep{2007VasquezJGRA, 2007PodestaApJ, 2010PodestaPhPl, 2012BorovskyJGRA}, we use $\Gamma = 3/2$, $N_{\parallel} = 2$ and $\rho_{\parallel} = 0.22$ as appropriate for Kraichnan turbulence \citep[see Table 1 of][]{2004CandiaJCAP}. We further note that there is evidence for velocity fluctuations in the solar wind following the Kraichnan spectrum while the magnetic field fluctuations follow the Kolmogorov's 5/3 scaling \citep{2013AlexandrovaSSRv,2010RobertsJGRA, 2009SalemApJ}. Therefore, we also briefly outline how our results change if we use parameters appropriate to Kolmogorov turbulence.


The anomalous electron diffusion timescale (in seconds) parallel to the magnetic field is given by
\begin{equation}
t_{\parallel} = \frac{D_{\parallel}}{v_{th_{e}}^2} \, \,  \, ,
\label{eq: nu_para}
\end{equation}
which can be used to determine the anomalous electron resistivity $\eta_{\parallel}$ defined as
\begin{equation}
\eta_{\parallel} = \frac{m_e}{t_{\parallel}\, n_{e} \, e^2} \, .
\label{eq: eta_para}
\end{equation}
The quantities $m_e$, $e$ and $n_e$ are electron mass in g, electron charge in esu and electron number density in cm$^{-3}$ (assumed to be the same as the proton number density $n_p$). The expression for the anomalous resistivity (Eq~\ref{eq: eta_para}) is in exact analogy with that for the collisional (Spitzer) resistivity $\eta_{coll}$, which is due to Coulomb collisions in an unmagnetized plasma:
\begin{equation}
\eta_{coll} = \frac{m_e}{t_{coll}\,n_e \, e^2} \, ,
\label{eq: eta_coll}
\end{equation} 
where the electron-electron Coulomb collisional timescale (in seconds) is \citep{2013NRLHuba},
\begin{equation}
t_{coll} = \biggl [ 2.91 \times 10^{-6} \, n_{e} \, {\rm ln} \Lambda  \,  T_{eV}^{-3/2} \biggr ]^{-1} \, .
\label{eq: nu_coll}
\end{equation}
\begin{figure}   
                                \centering
  \includegraphics[width=\hsize]{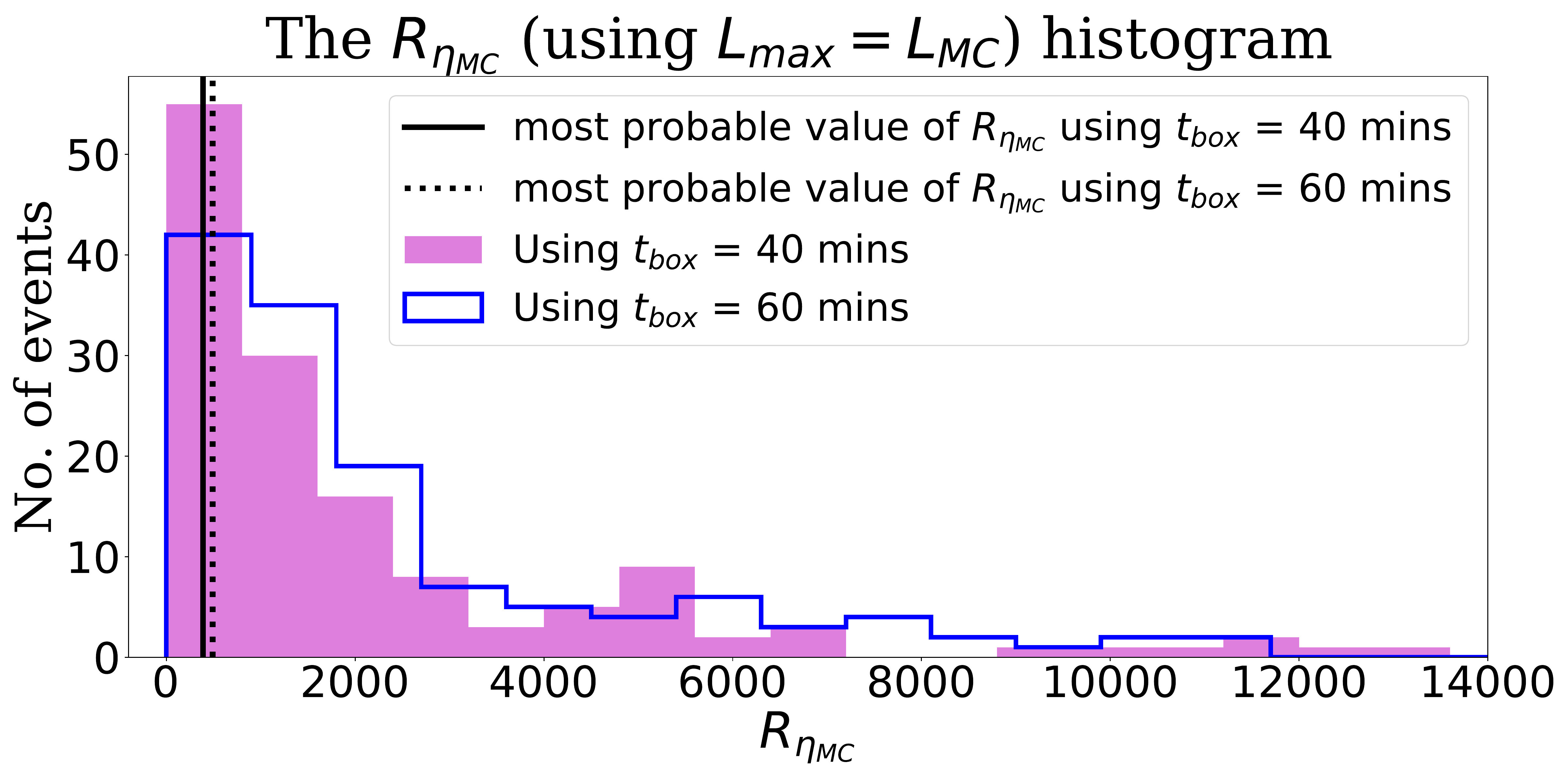}

\caption{Histograms of $R_{\eta_{MC}}$ (Equation~\ref{eq: resistivity_ratio_mc}) for all the MCs listed in Table~\ref{S - Table A} with $L_{max} = L_{MC}$ and parameters appropriate to the Kraichnan turbulence spectrum. The vertical lines indicate the most probable value. With $t_{box} = 40$ minutes the mean, median and most probable value of $R_{\eta_{MC}}$ are 7068, 1258 and 387 respectively. The corresponding values for $t_{box} = 60$ minutes are 10157, 1780 and 490 respectively.}
   \label{Figure - R_eta_mc}
   \end{figure}


The quantity ${\rm ln} \Lambda$ is the Coulomb logarithm (taken to be 20 for this study) and we take $T_{e} = 10 T_{p}$ \citep{2022DebeshSoPh}. Since we usually focus only on the (Spitzer) resistivity parallel to the background magnetic field \citep{2022DebeshSoPh}, we concentrate here only on $\eta_{\parallel}$. We can compare the anomalous resistivity with the collisional (Spitzer) resistivity using the following ratio
\begin{equation}
R_{\eta_{MC}} = \frac{\langle \eta_{\parallel} \rangle_{MC}}{\langle \eta_{coll} \rangle_{MC}}
\label{eq: resistivity_ratio_mc}
\end{equation}
where, as before, $\langle \, \rangle_{MC}$ denotes the temporal average inside the MC. The ratio $R_{\eta_{MC}}$ for all the MCs listed in Table~\ref{S - Table A} are shown in Figure~\ref{Figure - R_eta_mc}. The mean, median and most probable value are $7 \times 10^{3}$, $1.25 \times 10^{3}$ and 387 respectively for $t_{box} = 40$ minutes. For $t_{box} = 60$ minutes, they are $10^4$, $1.8 \times 10^{3}$ and 490 respectively. Using $L_{max} = 10^6$ km \citep{2009Prasad}, we find that the mean, median and most probable value of $R_{\eta_{MC}}$ are $3.7 \times 10^{4}$, $7.2 \times 10^{3}$ and $2 \times 10^{3}$ (for $t_{box} = 40$ minutes) respectively and $5.3 \times 10^{4}$, $10^{4}$ and 2770 (for $t_{box} = 60$ minutes) respectively.

We can similarly construct the ratio between the anomalous and Spitzer resistivity in the ambient solar wind background

\begin{equation}
R_{\eta_{BG}} = \frac{\langle \eta_{\parallel} \rangle_{BG}}{\langle \eta_{coll} \rangle_{BG}}
\label{eq: resistivity_ratio_bg}
\end{equation}
where $\langle \, \rangle_{BG}$ denotes the temporal average in the background. 
The mean, median and most probable value of $R_{\eta_{BG}}$ for $t_{box} = 40$ minutes are 1175, 0.68 and 0.04 respectively. For $t_{box} = 60$ minutes, the corresponding values are 1400, 0.88 and 0.06 respectively. With $L_{max} = 10^{6}$ km, the mean, median and most probable value of $R_{\eta_{BG}}$ are $5.7 \times 10^{3}$, 4.05 and 0.32 respectively for $t_{box} = 40$ minutes and 6900, 4.75 and 0.52 respectively for $t_{box}=60$ minutes. The mean values are biased by $\approx 22\%$ of the background solar wind periods associated with the 152 events listed in Table~\ref{S - Table A}.

To summarize, the anomalous resistivity arising out of turbulent fluctuations can be two to four orders of magnitude higher than the collisional (Spitzer) resistivity inside near-Earth MCs. On the other hand, the anomalous resistivity is generally lower than the collisional resistivity in the background solar wind. The sheath regions in front of ICMEs are known to be substantially more turbulent than the ICME interiors \cite{2021KilpuaFrASS}. 
Our preliminary calculations using sheath-appropriate values for $\delta B/B$ suggest that the anomalous resistivity can exceed the collisional one by as much as 4 to 5 orders of magnitude in the near-Earth sheath regions. This is an aspect that deserves further study. If we assume a Kolmogorov spectrum for the magnetic fluctuations (which amounts to using $\Gamma = 5/3$, $N_{\parallel} = 1.7$ and $\rho_{\parallel} = 0.2$) for the turbulent fluctuations, $R_{\eta_{MC}}$ and $R_{\eta_{BG}}$ are lower by a factor of $\approx$ 10. 


\section{Summary and Conclusions}
\label{S - conclusions}
Although solar wind turbulence is an extensively studied subject, not much attention has been devoted to turbulent fluctuations inside CMEs. We have surveyed the turbulence amplitudes of density and magnetic field fluctuations inside a well selected sample of near-Earth magnetic clouds using {\em in-situ} data from the Wind spacecraft. We divide the one minute resolution data into temporal boxes of duration $t_{box}$; we use $t_{box} = 40$ and 60 minutes. We compute the average density and magnetic field for each $t_{box}$ and compare it with the corresponding rms deviation to obtain the modulation indices of the proton density ($\delta n_{p}/n_{p}$) and total magnetic field ($\delta B/B$). We also compute the fluctuations of the acoustic Mach number in MCs. The magnetic field fluctuations scatter electrons, mimicking collisions. We estimate the anomalous resistivity arising out of such collisions. Our main results are as follows:
\begin{itemize}
\item
The most probable value of the average density modulation index inside near-Earth MCs ($\langle \delta n_{p}/n_{p} \rangle_{MC}$) ranges from 0.13 to 0.16 depending on the $t_{box}$ used, and is slightly larger than the density modulation index in the solar wind background ($\langle \delta n_{p}/n_{p} \rangle_{BG} \approx 0.08$ to 0.09 (depending on the $t_{box}$ used). On the other hand, the most probable value of the average magnetic field modulation index inside near-Earth MCs ($\langle \delta B/B \rangle_{MC}$) ranges from 0.04 to 0.05. The average magnetic field modulation index in the background solar wind ($\langle \delta B/B \rangle_{BG}$) is somewhat larger ($\approx 0.06$). We note that $\langle \delta B/B \rangle_{MC}$ is smaller than $\langle \delta n_{p}/n_{p} \rangle_{MC}$, suggesting that the magnetic fluctuations in MCs are probably associated with magnetosonic modes. We also find that the most probable value for fluctuations in the Mach number ($\delta M$) inside near-Earth MCs is $\approx 0.1$. The corresponding value inside the leading boundary layer of the ICME is around one order of magnitude smaller. In other words, although ICMEs propagate supersonically through the background solar wind, the turbulent fluctuations inside them are quite subsonic.
\item
The turbulent magnetic field fluctuations act as effective scattering centers for electrons. We calculate an anomalous resistivity arising from this process and find that it can be $\approx 500-1000$ times as large as the Spitzer resistivity inside near-Earth MCs. This will result in some enhanced Joule heating inside MCs. It is not clear if this level of enhanced Joule heating will substantially impact the overall energetics associated with CME propagation. 
CME sheaths, on the other hand, are characterized by much stronger turbulence, and a detailed study of such anomalous resistivity and attendant Joule heating in sheaths might yield interesting results that impact our understanding of CME energetics. Additional processes not considered here, such as reconnection (and thermal conduction from the reconnection sites) can also contribute to electron heating.
\end{itemize}

\section*{Acknowledgements}
\addcontentsline{toc}{section}{Acknowledgements}

DB acknowledges a PhD studentship from Indian Institute of Science Education and Research, Pune. We acknowledge a detailed review by
an anonymous referee that helped us improve this manuscript.




\bibliographystyle{mnras}
\bibliography{mnras4} 




\appendix

\begin{appendix} 
\section{Data Table}

\begin{landscape}

\begin{table}
  
\caption{
The list of the 152 Wind ICME events we use in this study. The arrival date and time of the ICME at the position of Wind measurement and the arrival and departure dates \& times of the associated magnetic clouds (MCs) are taken from Wind ICME catalogue (\url{https://wind.nasa.gov/ICMEindex.php}).Fr events indicate MCs with a single magnetic field rotation between $90^{\circ}$ and $180^{\circ}$ and F+ events indicate MCs with a single magnetic field rotation greater than $180^{\circ}$ \citep{18NCSo}. The 14 events marked with and asterisk (*) coincide with the near earth counterparts of 14 CMEs listed in \citet{17NSoPh}. }  
\label{S - Table A}

	\begin{center}
	
	\begin{tabular}{cccclccc}
		
		\hline
		\hline
		CME        & CME Arrival date & MC start  & MC end       & Flux rope \\
		event      & and time[UT]   &  date and   & date and     & type   \\
		number    &  (1AU)         &  time [UT]   &   time [UT]   &      \\
		\hline
		
		1    &  1995 03 04 , 00:36 & 1995 03 04 , 11:23 & 1995 03 05 , 03:06 & Fr \\
		2    &  1995 04 03 , 06:43 & 1995 04 03 , 12:45 & 1995 04 04 , 13:25 & F+ \\
		3    &  1995 06 30 , 09:21 & 1995 06 30 , 14:23 & 1995 07 02 , 16:47 & Fr \\
		4    &  1995 08 22 , 12:56 & 1995 08 22 , 22:19 & 1995 08 23 , 18:43 & Fr \\
		5    &  1995 09 26 , 15:57 & 1995 09 27 , 03:36 & 1995 09 27 , 21:21 & Fr \\
		6    &  1995 10 18 , 10:40 & 1995 10 18 , 19:11 & 1995 10 20 , 02:23 & Fr \\
		7    &  1996 02 15 , 15:07 & 1996 02 15 , 15:07 & 1996 02 16 , 08:59 & F+ \\
		8    &  1996 04 04 , 11:59 & 1996 04 04 , 11:59 & 1996 04 04 , 21:36 & Fr \\ 
		9    &  1996 05 16 , 22:47 & 1996 05 17 , 01:36 & 1996 05 17 , 11:58 & F+ \\
		10   &  1996 05 27 , 14:45 & 1996 05 27 , 14:45 & 1996 05 29 , 02:22 & Fr \\
		11   &  1996 07 01 , 13:05 & 1996 07 01 , 17:16 & 1996 07 02 , 10:17 & Fr \\
		12   &  1996 08 07 , 08:23 & 1996 08 07 , 11:59 & 1996 08 08 , 13:12 & Fr \\
		13   &  1996 12 24 , 01:26 & 1996 12 24 , 03:07 & 1996 12 25 , 11:44 & F+ \\ 
		14   &  1997 01 10 , 00:52 & 1997 01 10 , 04:47 & 1997 01 11 , 03:36 & F+ \\
		15   &  1997 04 10 , 17:02 & 1997 04 11 , 05:45 & 1997 04 11 , 19:10 & Fr \\
		16   &  1997 04 21 , 10:11 & 1997 04 21 , 11:59 & 1997 04 23 , 07:11 & F+ \\
		17   &  1997 05 15 , 01:15 & 1997 05 15 , 10:00 & 1997 05 16 , 02:37 & F+ \\
		18   &  1997 05 26 , 09:09 & 1997 05 26 , 15:35 & 1997 05 28 , 00:00 & Fr \\
		19   &  1997 06 08 , 15:43 & 1997 06 09 , 06:18 & 1997 06 09 , 23:01 & Fr \\
		20   &  1997 06 19 , 00:00 & 1997 06 19 , 05:31 & 1997 06 20 , 22:29 & Fr \\
		21   &  1997 07 15 , 03:10 & 1997 07 15 , 06:48 & 1997 07 16 , 11:16 & F+ \\
		22   &  1997 08 03 , 10:10 & 1997 08 03 , 13:55 & 1997 08 04 , 02:23 & Fr \\
		23   &  1997 08 17 , 01:56 & 1997 08 17 , 06:33 & 1997 08 17 , 20:09 & Fr \\
		24   &  1997 09 02 , 22:40 & 1997 09 03 , 08:38 & 1997 09 03 , 20:59 & Fr \\
		25   &  1997 09 18 , 00:30 & 1997 09 18 , 04:07 & 1997 09 19 , 23:59 & F+ \\
		26   &  1997 10 01 , 11:45 & 1997 10 01 , 17:08 & 1997 10 02 , 23:15 & Fr \\
		27   &  1997 10 10 , 03:08 & 1997 10 10 , 15:33 & 1997 10 11 , 22:00 & F+ \\
		28   &  1997 11 06 , 22:25 & 1997 11 07 , 06:00 & 1997 11 08 , 22:46 & F+ \\
		29   &  1997 11 22 , 09:12 & 1997 11 22 , 17:31 & 1997 11 23 , 18:43 & F+ \\
		30   &  1997 12 30 , 01:13 & 1997 12 30 , 09:35 & 1997 12 31 , 08:51 & Fr \\
		31   &  1998 01 06 , 13:29 & 1998 01 07 , 02:23 & 1998 01 08 , 07:54 & F+ \\
		32   &  1998 01 28 , 16:04 & 1998 01 29 , 13:12 & 1998 01 31 , 00:00 & F+ \\
		33   &  1998 03 25 , 10:48 & 1998 03 25 , 14:23 & 1998 03 26 , 08:57 & Fr \\
		34   &  1998 03 31 , 07:11 & 1998 03 31 , 11:59 & 1998 04 01 , 16:18 & Fr \\
		35   &  1998 05 01 , 21:21 & 1998 05 02 , 11:31 & 1998 05 03 , 16:47 & Fr \\
		36   &  1998 06 02 , 10:28 & 1998 06 02 , 10:28 & 1998 06 02 , 19:16 & Fr \\
		37   &  1998 06 24 , 10:47 & 1998 06 24 , 13:26 & 1998 06 25 , 22:33 & F+ \\
		38   &  1998 07 10 , 22:36 & 1998 07 10 , 22:36 & 1998 07 12 , 21:34 & F+ \\
		39   &  1998 08 19 , 18:40 & 1998 08 20 , 08:38 & 1998 08 21 , 20:09 & F+ \\
		40   &  1998 10 18 , 19:30 & 1998 10 19 , 04:19 & 1998 10 20 , 07:11 & F+ \\
		
		\hline
		
	\end{tabular}
\end{center}
\end{table}

\end{landscape}

\begin{landscape} 
\begin{table}
\caption{continued}
\begin{center}
	\begin{tabular}{cccclccc}
		\hline
		\hline
		CME        & CME Arrival date & MC start  & MC end       & Flux rope \\
		event      & and time[UT]   &  date and   & date and     & type   \\
		number    &  (1AU)         &  time [UT]   &   time [UT]   &      \\
		\hline
		
		41   &  1999 02 11 , 17:41 & 1999 02 11 , 17:41 & 1999 02 12 , 03:35 & Fr \\
		42   &  1999 07 02 , 00:27 & 1999 07 03 , 08:09 & 1999 07 05 , 13:13 & Fr \\
		43   &  1999 09 21 , 18:57 & 1999 09 21 , 18:57 & 1999 09 22 , 11:31 & Fr \\
		44   &  2000 02 11 , 23:34 & 2000 02 12 , 12:20 & 2000 02 13 , 00:35 & Fr \\
		45   &  2000 02 20 , 21:03 & 2000 02 21 , 14:24 & 2000 02 22 , 13:16 & Fr \\

		46   &  2000 03 01 , 01:58 & 2000 03 01 , 03:21 & 2000 03 02 , 03:07 & Fr \\
		47   &  2000 07 01 , 07:12 & 2000 07 01 , 07:12 & 2000 07 02 , 03:34 & Fr \\
		48   &  2000 07 11 , 22:35 & 2000 07 11 , 22:35 & 2000 07 13 , 04:33 & Fr \\
		49   &  2000 07 28 , 06:38 & 2000 07 28 , 14:24 & 2000 07 29 , 10:06 & F+ \\
		50   &  2000 09 02 , 23:16 & 2000 09 02 , 23:16 & 2000 09 03 , 22:32 & Fr \\
		51   &  2000 10 03 , 01:02 & 2000 10 03 , 09:36 & 2000 10 05 , 03:34 & F+ \\
		52   &  2000 10 12 , 22:33 & 2000 10 13 , 18:24 & 2000 10 14 , 19:12 & Fr \\
		53   &  2000 11 06 , 09:30 & 2000 11 06 , 23:05 & 2000 11 07 , 18:05 & Fr \\
		54   &  2000 11 26 , 11:43 & 2000 11 27 , 09:30 & 2000 11 28 , 09:36 & Fr \\
		55   &  2001 04 21 , 15:29 & 2001 04 22 , 00:28 & 2001 04 23 , 01:11 & Fr \\
		56   &  2001 10 21 , 16:39 & 2001 10 22 , 01:17 & 2001 10 23 , 00:47 & Fr \\
		57   &  2001 11 24 , 05:51 & 2001 11 24 , 15:47 & 2001 11 25 , 13:17 & Fr \\
		58   &  2001 12 29 , 05:16 & 2001 12 30 , 03:24 & 2001 12 30 , 19:10 & Fr \\
		59   &  2002 02 28 , 05:06 & 2002 02 28 , 19:11 & 2002 03 01 , 23:15 & Fr \\
		60   &  2002 03 18 , 13:14 & 2002 03 19 , 06:14 & 2002 03 20 , 15:36 & Fr \\
		61   &  2002 03 23 , 11:24 & 2002 03 24 , 13:11 & 2002 03 25 , 21:36 & Fr \\
		62   &  2002 04 17 , 11:01 & 2002 04 17 , 21:36 & 2002 04 19 , 08:22 & F+ \\
		63   &  2002 07 17 , 15:56 & 2002 07 18 , 13:26 & 2002 07 19 , 09:35 & Fr \\
		64   &  2002 08 18 , 18:40 & 2002 08 19 , 19:12 & 2002 08 21 , 13:25 & Fr \\
		65   &  2002 08 26 , 11:16 & 2002 08 26 , 14:23 & 2002 08 27 , 10:47 & Fr \\
		66   &  2002 09 30 , 07:54 & 2002 09 30 , 22:04 & 2002 10 01 , 20:08 & F+ \\
		67   &  2002 12 21 , 03:21 & 2002 12 21 , 10:20 & 2002 12 22 , 15:36 & Fr \\
		68   &  2003 01 26 , 21:43 & 2003 01 27 , 01:40 & 2003 01 27 , 16:04 & Fr \\
		69   &  2003 02 01 , 13:06 & 2003 02 02 , 19:11 & 2003 02 03 , 09:35 & Fr \\
		70   &  2003 03 20 , 04:30 & 2003 03 20 , 11:54 & 2003 03 20 , 22:22 & Fr \\
		71   &  2003 06 16 , 22:33 & 2003 06 17 , 17:48 & 2003 06 18 , 08:18 & Fr \\
		72   &  2003 08 04 , 20:23 & 2003 08 05 , 01:10 & 2003 08 06 , 02:23 & Fr \\
		73   &  2003 11 20 , 08:35 & 2003 11 20 , 11:31 & 2003 11 21 , 01:40 & Fr \\
		74   &  2004 04 03 , 09:55 & 2004 04 04 , 01:11 & 2004 04 05 , 19:11 & F+ \\
		75   &  2004 09 17 , 20:52 & 2004 09 18 , 12:28 & 2004 09 19 , 16:58 & Fr \\
		
		76   &  2005 05 15 , 02:10 & 2005 05 15 , 05:31 & 2005 05 16 , 22:47 & F+ \\
		77   &  2005 05 20 , 04:47 & 2005 05 20 , 09:35 & 2005 05 22 , 02:23 & F+ \\
		78   &  2005 07 17 , 14:52 & 2005 07 17 , 14:52 & 2005 07 18 , 05:59 & Fr \\
		79   &  2005 10 31 , 02:23 & 2005 10 31 , 02:23 & 2005 10 31 , 18:42 & Fr \\
		80   &  2006 02 05 , 18:14 & 2006 02 05 , 20:23 & 2006 02 06 , 11:59 & F+ \\
		81   &  2006 09 30 , 02:52 & 2006 09 30 , 08:23 & 2006 09 30 , 22:03 & F+ \\
		82   &  2006 11 18 , 07:11 & 2006 11 18 , 07:11 & 2006 11 20 , 04:47 & Fr \\
		83   &  2007 05 21 , 22:40 & 2007 05 21 , 22:45 & 2007 05 22 , 13:25 & Fr \\
		84   &  2007 06 08 , 05:45 & 2007 06 08 , 05:45 & 2007 06 09 , 05:15 & Fr \\
		85   &  2007 11 19 , 17:22 & 2007 11 20 , 00:33 & 2007 11 20 , 11:31 & Fr \\
		
		\hline
		
	\end{tabular}
\end{center}
\end{table}
\end{landscape}

\begin{landscape} 

\begin{table}
\caption{continued}
\begin{center}
	\begin{tabular}{cccclccc}
		\hline
		\hline
		CME        & CME Arrival date & MC start  & MC end       & Flux rope \\
		event      & and time[UT]   &  date and   & date and     & type   \\
		number    &  (1AU)         &  time [UT]   &   time [UT]   &      \\
		\hline
		
		86   &  2008 05 23 , 01:12 & 2008 05 23 , 01:12 & 2008 05 23 , 10:46 & F+ \\
		87   &  2008 09 03 , 16:33 & 2008 09 03 , 16:33 & 2008 09 04 , 03:49 & F+ \\
		88   &  2008 09 17 , 00:43 & 2008 09 17 , 03:57 & 2008 09 18 , 08:09 & Fr \\
		89   &  2008 12 04 , 11:59 & 2008 12 04 , 16:47 & 2008 12 05 , 10:47 & Fr \\
		90   &  2008 12 17 , 03:35 & 2008 12 17 , 03:35 & 2008 12 17 , 15:35 & Fr \\
		91   &  2009 02 03 , 19:21 & 2009 02 04 , 01:12 & 2009 02 04 , 19:40 & F+ \\
		92   &  2009 03 11 , 22:04 & 2009 03 12 , 01:12 & 2009 03 13 , 01:40 & F+ \\
		93   &  2009 04 22 , 11:16 & 2009 04 22 , 14:09 & 2009 04 22 , 20:37 & Fr \\
		94   &  2009 06 03 , 13:40 & 2009 06 03 , 20:52 & 2009 06 05 , 05:31 & Fr \\
		95   &  2009 06 27 , 11:02 & 2009 06 27 , 17:59 & 2009 06 28 , 20:24 & F+ \\

		96   &  2009 07 21 , 02:53 & 2009 07 21 , 04:48 & 2009 07 22 , 03:36 & Fr \\
		97   &  2009 09 10 , 10:19 & 2009 09 10 , 10:19 & 2009 09 10 , 19:26 & Fr \\
		
		98   &  2009 09 30 , 00:44 & 2009 09 30 , 06:59 & 2009 09 30 , 19:11 & Fr \\
		99   &  2009 10 29 , 01:26 & 2009 10 29 , 01:26 & 2009 10 29 , 23:45 & F+ \\
		100   &  2009 11 14 , 10:47 & 2009 11 14 , 10:47 & 2009 11 15 , 11:45 & Fr \\
		101   &  2009 12 12 , 04:47 & 2009 12 12 , 19:26 & 2009 12 14 , 04:47 & Fr \\
		102   &  2010 01 01 , 22:04 & 2010 01 02 , 00:14 & 2010 01 03 , 09:06 & Fr \\
		103   &  2010 02 07 , 18:04 & 2010 02 07 , 19:11 & 2010 02 09 , 05:42 & Fr \\
		104*   &  2010 03 23 , 22:29 & 2010 03 23 , 22:33 & 2010 03 24 , 15:36 & Fr \\
		105*   &  2010 04 05 , 07:55 & 2010 04 05 , 11:59 & 2010 04 06 , 16:48 & Fr \\
		106*   &  2010 04 11 , 12:20 & 2010 04 11 , 21:36 & 2010 04 12 , 14:12 & Fr \\
		107   &  2010 05 28 , 01:55 & 2010 05 28 , 19:12 & 2010 05 29 , 17:58 & Fr \\
		108*   &  2010 06 21 , 03:35 & 2010 06 21 , 06:28 & 2010 06 22 , 12:43 & Fr \\
		109*   &  2010 09 15 , 02:24 & 2010 09 15 , 02:24 & 2010 09 16 , 11:58 & Fr \\
		110*   &  2010 10 31 , 02:09 & 2010 10 31 , 05:16 & 2010 11 01 , 20:38 & Fr \\
		111   &  2010 12 19 , 20:35 & 2010 12 19 , 22:33 & 2010 12 20 , 22:04 & F+ \\
		112   &  2011 01 24 , 06:43 & 2011 01 24 , 10:33 & 2011 01 25 , 22:04 & F+ \\
		113*   &  2011 03 29 , 15:12 & 2011 03 29 , 23:59 & 2011 04 01 , 14:52 & Fr \\
		114   &  2011 05 28 , 00:14 & 2011 05 28 , 05:31 & 2011 05 28 , 22:47 & F+ \\
		115   &  2011 06 04 , 20:06 & 2011 06 05 , 01:12 & 2011 06 05 , 18:13 & Fr \\
		116   &  2011 07 03 , 19:12 & 2011 07 03 , 19:12 & 2011 07 04 , 19:12 & Fr \\
		117*   &  2011 09 17 , 02:57 & 2011 09 17 , 15:35 & 2011 09 18 , 21:07 & Fr \\
		118   &  2012 02 14 , 07:11 & 2012 02 14 , 20:52 & 2012 02 16 , 04:47 & Fr \\
		119   &  2012 04 05 , 14:23 & 2012 04 05 , 19:41 & 2012 04 06 , 21:36 & Fr \\
		120   &  2012 05 03 , 00:59 & 2012 05 04 , 03:36 & 2012 05 05 , 11:22 & Fr \\
		121   &  2012 05 16 , 12:28 & 2012 05 16 , 16:04 & 2012 05 18 , 02:11 & Fr \\
		122   &  2012 06 11 , 02:52 & 2012 06 11 , 11:31 & 2012 06 12 , 05:16 & Fr \\
		123*   &  2012 06 16 , 09:03 & 2012 06 16 , 22:01 & 2012 06 17 , 11:23 & F+ \\
		124*   &  2012 07 14 , 17:39 & 2012 07 15 , 06:14 & 2012 07 17 , 03:22 & Fr \\
		125   &  2012 08 12 , 12:37 & 2012 08 12 , 19:12 & 2012 08 13 , 05:01 & Fr \\
		126   &  2012 08 18 , 03:25 & 2012 08 18 , 19:12 & 2012 08 19 , 08:22 & Fr \\
		127*   &  2012 10 08 , 04:12 & 2012 10 08 , 15:50 & 2012 10 09 , 17:17 & Fr \\
		128   &  2012 10 12 , 08:09 & 2012 10 12 , 18:29 & 2012 10 13 , 09:14 & Fr \\
		129*   &  2012 10 31 , 14:28 & 2012 10 31 , 23:35 & 2012 11 02 , 05:21 & F+ \\
		130*   &  2013 03 17 , 05:21 & 2013 03 17 , 14:09 & 2013 03 19 , 16:04 & Fr \\

		\hline
		
	\end{tabular}
\end{center}
\end{table}

\end{landscape}

\begin{landscape} 
\begin{table}
\caption{continued}
\begin{center}
	\begin{tabular}{cccclccc}
		\hline
		\hline
		CME        & CME Arrival date & MC start  & MC end       & Flux rope \\
		event      & and time[UT]   &  date and   & date and     & type   \\
		number    &  (1AU)         &  time [UT]   &   time [UT]   &      \\
		\hline

		
		131*   &  2013 04 13 , 22:13 & 2013 04 14 , 17:02 & 2013 04 17 , 05:30 & F+ \\
		132   &  2013 04 30 , 08:52 & 2013 04 30 , 12:00 & 2013 05 01 , 07:12 & Fr \\
		133   &  2013 05 14 , 02:23 & 2013 05 14 , 06:00 & 2013 05 15 , 06:28 & Fr \\
		134   &  2013 06 06 , 02:09 & 2013 06 06 , 14:23 & 2013 06 08 , 00:00 & F+ \\
		135   &  2013 06 27 , 13:51 & 2013 06 28 , 02:23 & 2013 06 29 , 11:59 & Fr \\
		136   &  2013 09 01 , 06:14 & 2013 09 01 , 13:55 & 2013 09 02 , 01:56 & Fr \\
		137   &  2013 10 30 , 18:14 & 2013 10 30 , 18:14 & 2013 10 31 , 05:30 & Fr \\
		138   &  2013 11 08 , 21:07 & 2013 11 08 , 23:59 & 2013 11 09 , 06:14 & Fr \\
		139   &  2013 11 23 , 00:14 & 2013 11 23 , 04:47 & 2013 11 23 , 15:35 & Fr \\
		140   &  2013 12 14 , 16:47 & 2013 12 15 , 16:47 & 2013 12 16 , 05:30 & Fr \\

		141   &  2013 12 24 , 20:36 & 2013 12 25 , 04:47 & 2013 12 25 , 17:59 & F+ \\
		142   &  2014 04 05 , 09:58 & 2014 04 05 , 22:18 & 2014 04 07 , 14:24 & Fr \\
		143   &  2014 04 11 , 06:57 & 2014 04 11 , 06:57 & 2014 04 12 , 20:52 & F+ \\
		144   &  2014 04 20 , 10:20 & 2014 04 21 , 07:41 & 2014 04 22 , 06:12 & Fr \\
		145   &  2014 04 29 , 19:11 & 2014 04 29 , 19:11 & 2014 04 30 , 16:33 & Fr \\
		146   &  2014 06 29 , 16:47 & 2014 06 29 , 20:53 & 2014 06 30 , 11:15 & Fr \\
		147   &  2014 08 19 , 05:49 & 2014 08 19 , 17:59 & 2014 08 21 , 19:09 & F+ \\
		148   &  2014 08 26 , 02:40 & 2014 08 27 , 03:07 & 2014 08 27 , 21:49 & Fr \\
		149   &  2015 01 07 , 05:38 & 2015 01 07 , 06:28 & 2015 01 07 , 21:07 & F+ \\
		150   &  2015 09 07 , 13:05 & 2015 09 07 , 23:31 & 2015 09 09 , 14:52 & F+ \\
		151   &  2015 10 06 , 21:35 & 2015 10 06 , 21:35 & 2015 10 07 , 10:03 & Fr \\
		152   &  2015 12 19 , 15:35 & 2015 12 20 , 13:40 & 2015 12 21 , 23:02 & Fr \\
		\hline

  \end{tabular}
  \end{center}
\end{table}   
\end{landscape}   

\begin{landscape} 
	\newpage
	\begin{table}
		\begin{center}
	\caption{This table displays the statistical outputs (mean, median and most probable value) related to the proton density, total magnetic field, acoustic Mach number fluctuations and the resistivity ratio inside the MC and in the background. $\langle \delta M \rangle_{fB}$ and $\langle \delta M \rangle_{tB}$ represent the acoustic Mach number fluctuations in the MC front and trailing boundary layers respectively.}
	\label{S - Table B}
	\label{tab:anysymbols}
	\begin{tabular}{ccccc}
		\hline
		Parameter & $t_{box}$ (minute) & Mean & Median & Most Probable Value\\
		\hline
		\hline
		$\langle \delta n_{p}/n_{p} \rangle_{MC}$ & 40 & 0.24 & 0.20 & 0.13 \\
		$\langle \delta n_{p}/n_{p} \rangle_{MC}$ & 60 & 0.28 & 0.23 & 0.16 \\
		\hline
		$\langle \delta n_{p}/n_{p} \rangle_{BG}$ & 40 & 0.11 & 0.10 & 0.08 \\
		$\langle \delta n_{p}/n_{p} \rangle_{BG}$ & 60 & 0.12 & 0.11 & 0.09 \\
		\hline
		\hline
		$\langle \delta B/B \rangle_{MC}$ & 40 & 0.044 & 0.042 & 0.038 \\
		$\langle \delta B/B \rangle_{MC}$ & 60 & 0.056 & 0.052 & 0.050 \\
		\hline
		$\langle \delta B/B \rangle_{BG}$ & 40 & 0.067 & 0.060 & 0.059 \\
		$\langle \delta B/B \rangle_{BG}$ & 60 & 0.074 & 0.068 & 0.063 \\
		\hline
		\hline
		[using $\gamma = 5/3$] & & & & \\
		$\langle \delta M \rangle_{MC}$ & 40 & 0.15 & 0.12 & 0.09 \\
		$\langle \delta M \rangle_{MC}$ & 60 & 0.17 & 0.14 & 0.10 \\
		\hline
		$\langle \delta M \rangle_{BG}$ & 40 & 0.84 & 0.46 & 0.12 \\
		$\langle \delta M \rangle_{BG}$ & 60 & 0.95 & 0.52 & 0.14 \\
		\hline
		$\langle \delta M \rangle_{fB}$ & 10 & 0.14 & 0.07 & 0.05 \\
		$\langle \delta M \rangle_{tB}$ & 10 & 0.16 & 0.05 & 0.04 \\
		\hline
		[using $\gamma = 1.2$] & & & & \\
		$\langle \delta M \rangle_{MC}$ & 40 & 0.17 & 0.14 & 0.11 \\
		$\langle \delta M \rangle_{MC}$ & 60 & 0.21 & 0.16 & 0.12 \\
		\hline
		$\langle \delta M \rangle_{BG}$ & 40 & 0.98 & 0.54 & 0.14 \\
		$\langle \delta M \rangle_{BG}$ & 60 & 1.09 & 0.60 & 0.15 \\
		\hline
		$\langle \delta M \rangle_{fB}$ & 10 & 0.17 & 0.08 & 0.05 \\
		$\langle \delta M \rangle_{tB}$ & 10 & 0.18 & 0.06 & 0.04 \\
		\hline
		\hline
		[using $L_{max} = L_{MC}$] & & & & \\
		$ R_{\eta_{MC}}$ & 40 & 7068.20 & 1258.70 & 387.10 \\
		$ R_{\eta_{MC}}$ & 60 & 10157.41 & 1780 & 490.52 \\
		\hline
		$ R_{\eta_{BG}}$ & 40 & 1175 & 0.68 & 0.04 \\
		$ R_{\eta_{BG}}$ & 60 & 1398.40 & 0.88 & 0.06 \\
		\hline
		[using $L_{max} = 10^6$km] & & & & \\
		$ R_{\eta_{MC}}$ & 40 & 37344.12 & 7218.51 & 2025.50 \\
		$ R_{\eta_{MC}}$ & 60 & 53545.63 & 10809.51 & 2770.22 \\
		\hline
		$ R_{\eta_{BG}}$ & 40 & 5777.11 & 4.05 & 0.32 \\
		$ R_{\eta_{BG}}$ & 60 & 6898.70 & 4.75 & 0.52 \\
		\hline
		\hline
		
	\end{tabular}
\end{center}
\end{table}   
\end{landscape} 

\end{appendix}
%

\bsp	
\label{lastpage}
\end{document}